\documentclass[twocolumn,showpacs,preprintnumbers,superscriptaddress]{revtex4}

\usepackage{amsmath,amssymb,graphicx,bm}

\begin{document}

\title{Restrictions on purely kinetic k-essence}

\author{Rong-Jia Yang}
\email{yangrj08@gmail.com}
 \affiliation{College of Physical Science and Technology, Hebei University, Baoding 071002, China}

\author{Xiang-Ting, Gao}
 \affiliation{College of Physical Science and Technology, Hebei University, Baoding 071002, China}

\date{\today}

\begin{abstract}
We restrict purely kinetic k-essence. Assuming the equation of state is a power law of the kinetic energy: $w=w_0X^{\alpha}$, to obtain accelerated phases, we must have $\alpha>0$ as one of necessary conditions, constrained from the conditions for stability and causality, and the k-essence must behave like phantom. We also study the evolutions of the equation of state and the speed of sound with numerical simulation.

\end{abstract}

\maketitle

\section{Introduction}

In the last decade a convergence of independent cosmological
observations suggested that the Universe is experiencing accelerated
expansion. An unknown energy component, dubbed as dark energy, is
proposed to explain this acceleration. Dark energy almost equally
distributes in the Universe, and its pressure is negative. The simplest and most theoretically appealing candidate of
dark energy is the vacuum energy (or the cosmological constant
$\Lambda$) with a constant equation of state (EoS) parameter $w=-1$.
This scenario is in general agreement with the current astronomical
observations, but has difficulties to reconcile the small
observational value of dark energy density with estimates from
quantum field theories; this is the cosmological constant problem.
Recently it was shown that $\Lambda$CDM model may also suffer age problem \cite{Yang2010}.
It is thus natural to pursue alternative possibilities to explain
the mystery of dark energy. Over the past decade numerous dark energy models have been proposed,
such as quintessence, phantom, k-essence, tachyon, (Generalized) Chaplygin Gas, DGP, etc. k-essence, a simple approach toward constructing a model for an accelerated expansion of the Universe, is to work with the idea that the unknown dark energy component is due exclusively
to a minimally coupled scalar field $\phi$ with non-canonical kinetic energy which results in the negative pressure \cite{Arm00}. A feature of k-essence models is that the negative pressure results from the non-linear kinetic energy of the
scalar field. Secondly, because of the dynamical attractor behavior,
cosmic evolution is insensitive to initial conditions in k-essence
theories. Thirdly, k-essence changes its speed of evolution in
dynamic response to changes in the background EoS.

K-essence scenario has received much attention,  it was originally proposed as a model for inflation
\cite{Damour1999}, and then as a model for dark energy \cite{Arm00}.
In several cases, k-essence cannot be observationally distinguished from quintessence \cite{Malquarti2003}.
A method to obtain a phantom version of FRW k-essence cosmologies was devised in \cite{Aguirregabiria2004}.
The stability of k-essence was studied in \cite{Abramo2006}.
Dynamics of k-essence were discussed in \cite{Rendall2006}.
Conditions for stable tracker solutions for k-essence in a general cosmological
background were derived in \cite{Das2006}.
Slow-roll conditions for thawing k-essence were obtained in \cite{Chiba2009}.
A connection between the holographic dark energy density and the kinetic k-essence energy density
was discussed in \cite{Cruz2009}.
An holographic k-essence model of dark energy was proposed in \cite{Granda2009}.
The geometrical diagnostic for purely kinetic k-essence dark energy was discussed in \cite{Gao2010}.
The equivalence of a barotropic perfect fluid with a k-essence scalar field was considered in \cite{Arroja2010}.
A linear k-essence field model on a brane universe was examined in \cite{Chimento2009}.
Models of dark energy with purely kinetic multiple k-essence sources that allow for the crossing
of the phantom divide line were investigated in \cite{Sur2009}.
The thermodynamic properties of of k-essence was discussed in \cite{Bilic2008}.
Models of k-essence unified dark matter were discussed in \cite{Daniele2007,Scherrer2004,Bose2009}.
Theoretical and observational Constraints on k-essence dark energy models were discussed in \cite{Yang2009,Yang2008,Yang2008a}.
In Ref. \cite{Sen2006}, a model independent method of reconstructing the Lagrangian for the k-essence field by using three parametrizations
for the Hubble parameter $H(z)$ was studied in detail. With assumptions on the EoS of k-essence as functions of the scale factor $a$, Ref. \cite{Putter2007}
discussed the forms of the Lagrangians. In this paper, we will restrict on purely kinetic k-essence with some assumptions on the EoS of k-essence
as functions of the kinetic energy $X$, and study the evolution of purely kinetic k-essence.

This paper is organized as follows, in the following section, we review the model of k-essence and study its evolution. In Sec. III, we restrict on purely kinetic k-essence. Finally, we shall close with a few concluding remarks in Sec. IV.

\section{Briefly Review on k-essence}
As a candidate of dark energy,
k-essence is defined as a scalar field $\phi$ with
non-linear kinetic terms which appears generically in the effective
action in string and supergravity theories, its action minimally
coupled with gravity generically may be expressed as \cite{Arm00, Damour1999, Garriga1999}
\begin{eqnarray}
S_{\phi}=\int d^4x\sqrt{-g}\left[-\frac{R}{2}+p(\phi,X)\right],
\end{eqnarray}
where $X\equiv\frac{1}{2}\partial_{\mu}\phi\partial^{\mu}\phi$. We assume a flat and homogeneous Friedmann-Robertson-Walker (FRW) space-time
and work in units $8\pi G=c=1$. In this case, we have $X=\frac{1}{2}\dot{\phi}^2$, implying $X \geq 0$.

The Lagrangian $p$ and the energy density of k-essence take the forms, respectively:
\begin{eqnarray}
p&=&V(\phi)F(X),\\
\rho &=&V(\phi)[2XF_{X}-F],
\end{eqnarray}
here $F(X)$ is a function of
the kinetic energy $X$ and $F_{X}\equiv dF/dX$. The corresponding EoS parameter and the effective sound speed are
given by

\begin{eqnarray}
\label{w}w&=&\frac{F}{2XF_{X}-F}, \\
\label{c}c^{2}_{\rm s}&=&\frac{\partial p/\partial
X}{\partial\rho /\partial X}=\frac{F_{X}}{F_{X}+2XF_{XX}},
\end{eqnarray}
with $F_{XX}\equiv d^{2}F/dX^{2}$. The definition of the sound speed
comes from the equation describing the evolution of linear
adiabatic perturbations in a k-essence dominated Universe \cite{Garriga1999} (the non-adiabatic perturbation was discussed in \cite{Unnikrishnan2010}, here we
only consider the case of adiabatic perturbations). Perturbations can become unstable if the sound speed is imaginary, $c_{\rm s}^2<0$, so we insist on $c_{\rm s}^2 > 0$. Another potentially interesting requirement to consider is $c_{\rm s}^2 \leq 1$, which says that the sound speed should not exceed the speed of light, which suggests violation of causality. Though this is an open problem (see e. g. \cite{Babichev2008,Bruneton2007,Kang2007,Bonvin2006,Gorini2008,Ellis2007}), we still impose this condition.

Note that the EoS $w$ and the sound speed $c^{2}_{\rm s}$ do not depend explicitly on $V(\phi)$ in any
case. Without loss of generality, we take $V(\phi)$ to be a constant
discussed in Refs. \cite{Scherrer2004,Sen2006,Putter2007,Yang2009,Yang2008,Yang2008a,Chim04}; in other words, we consider a purely
kinetic $k$-essence models in which $p=V_0F(X)$. In Refs.
\cite{Scherrer2004,Chim04,Yang2008a}, a theoretical constraint on purely kinetic
k-essence was obtained
\begin{eqnarray}
\label{5}XF^2_X=k_0a^{-6},
\end{eqnarray}
where $k_0$ is a constant of integration. Given any forms of $F(X)$,
Eq. (\ref{5}) gives the evolution of $X$, then the evolution of
the EoS parameter $w$ and the sound speed $c^{2}_{\rm s}$ as a function of the scale factor $a$. In this case, solution
(\ref{5}) can be considered as a theoretical constraint on purely kinetic k-essence. Next, we constrain purely kinetic k-essence
by using Eq. (\ref{5}) with assumptions on the EoS $w$.

\section{Restrictions on purely kinetic k-essence}
We here restrict on the evolution of the k-essence by using Eq. (\ref{5}) with a particular ansatzes for the EoS of k-essence. From Eq. (\ref{w}),
we see the EoS $w$ depend explicitly on the kinetic energy $X$, while depend implicitly on $a$ by Eq. (\ref{5}), so we assume the EoS is a function of the
kinetic energy $X$.

As a simple case, we consider a power law: $w=w_0X^{\alpha}$ with $w_0$ a nonzero constant. When $\alpha=0$, $w$ is a constant. It is will known that in this case the energy density is $\rho\propto a^{-3(1+w)}$, for radiation $w=1/3$, matter $w = 0$, and a cosmological constant $w = -1$. For a nonzero constant $w$, we obtain from Eq. (\ref{w})
\begin{eqnarray}
\label{F1}
F(X)=F_1X^{\frac{1+w}{2w}},
\end{eqnarray}
where $F_1$ is a nonzero constant of integration. To obtain this equation, we do not assume a constant potential. So $F_1$ can be a function of $\phi$ and
takes the role of potential. 

Combining Eqs. (\ref{5}) and (\ref{F1}), we have
\begin{eqnarray}
X=k_1a^{-6w}.
\end{eqnarray}
This is the evolution of the kinetic energy $X$. This equation can also be obtain from the fact $F(X)=w\rho \propto a^{-3(1+w)}$. We note that when the k-essence evolves as radiation ($w=1/3$), the kinetic energy $X$ evolves as
spatial curvature. Combining Eqs. (\ref{c}) and (\ref{F2}), we get
\begin{eqnarray}
c^{2}_{\rm s}=w.
\end{eqnarray}
Considering the condition for stability, we have $w\geq0$, meaning there is no accelerated phase in purely kinetic k-essence model with constant EoS.

When $\alpha\neq 0$, we obtain from Eq. (\ref{w})
\begin{eqnarray}
\label{F2}
F(X)=F_2\sqrt{X}e^{-\beta X^{-\alpha}},
\end{eqnarray}
where $\beta=1/(2\alpha w_0)$ and $F_2$ a nonzero constant of integration. Also $F_2$ can be a function of $\phi$ and
takes the role of potential. Considering the case of a constant potential, and combining Eqs. (\ref{5}) and (\ref{F2}), we obtain
\begin{eqnarray}
\label{r}
(1+2\alpha\beta X^{-\alpha})^2e^{-2\beta X^{-\alpha}}=k_2a^{-6},
\end{eqnarray}
where $k_2$ is a positive constant. This is the equation restricting the evolution of k-essence. Combining Eqs. (\ref{c}) and (\ref{F2}), we get
\begin{eqnarray}
c^{2}_{\rm s}=\frac{(2\alpha\beta+X^{\alpha})X^{\alpha}}{2\alpha\beta[2\alpha\beta+(1-2\alpha)X^{2\alpha\beta}]}.
\end{eqnarray}
From the conditions of stability and causality: $0<c^{2}_{\rm s}\leq 1$, we obtain:

(I) $\left(-\alpha-\sqrt{\alpha^2+1}\right)X^{-\alpha}\leq w_0<-X^{-\alpha}$ (there are accelerated phases) or $0<w_0\leq \left(-\alpha+\sqrt{\alpha^2+1}\right)X^{-\alpha}$ (there are no accelerated phases) for $\alpha>0$;

(II) $0<w_0\leq \left(-\alpha+\sqrt{\alpha^2+1}\right)X^{-\alpha}$ (there are no accelerated phases) for $\alpha<0$.

These conditions constrain on not only $\alpha$ and $w_0$, but also $X$. It is obvious that to obtain accelerated phases we must have $\alpha>0$
as one of necessary conditions, and k-essence evolves as phantom. 

We study the evolution of k-essence with numerical simulation by using Eq. (\ref{r}). We only concentrate on the cases in which there are accelerated phases.
In Fig. 1, we plot the evolution of EoS of k-essence as the function of $a$ for some value of $\alpha$, $w_0$, $k_2$, and $X$.
It is obvious that in all these cases the EoS $w$ increases with $a$, evolves like phantom with $w<-1$, and run close to cosmological constant in the future.
In Fig. 2, we plot the evolution of the speed of sound $c^{2}_{\rm s}$ as the function of $a$ for the same value of $\alpha$, $w_0$, $k_2$,
and $X$ taken in figures of $w(a)$ in the same line. It is obvious that in all these cases the speed of sound $c^{2}_{\rm s}$ decreases with $a$,
and $0<c^{2}_{\rm s}\leq 1$ which means the model is stability and causality.

\begin{figure}
\includegraphics[width=8cm]{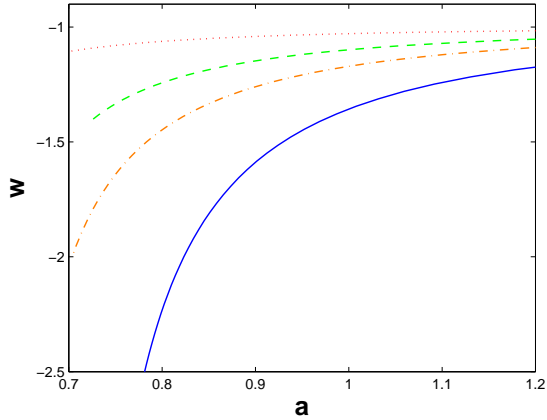}
\caption{The evolution of EoS $w$ as the function of $a$, for dot line, $\alpha=0.2$, $w_0=-1$, $k_2=0.1$, and $1< X< 2$; for dash line, 
$\alpha=0.5$, $w_0=-1$, $k_2=0.05$, and $1< X< 2$; for dash-dot line, $\alpha=1$, $w_0=-1$, $k_2=0.05$, and $1< X< 2$; for solid line $\alpha=2$, $w_0=-2$, $k_2=0.1$, and $0.75< X< 1.25$.}
\end{figure}

\begin{figure}
\includegraphics[width=8cm]{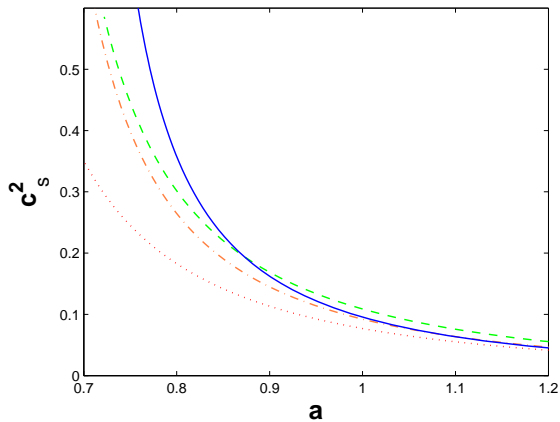}
\caption{The evolution of the speed of sound $c^{2}_{\rm s}$ as the function of $a$ for the same value of $\alpha$, $w_0$, $k$, and $X$ taken in Fig. 1.}
\end{figure}

\section{Conclusions and discussions}
With a assumption that the EoS is a power law of the
kinetic energy: $w=w_0X^{\alpha}$, we have restricted on purely kinetic k-essence and studied its evolution by using the theoretical constraint, Eq. (\ref{5}).
We have restricted the forms of the Lagrangian $p$. From the conditions for stability and causality, we have Constrained on $\alpha$, $w_0$ and $X$. When $\alpha\leq0$, there are no accelerated phases. To obtain accelerated phases, we must have $\alpha>0$ as well as $w_0<0$ (this is obvious) as necessary conditions, and the k-essence behaves like phantom. We have plotted the evolutions of EoS and the speed of sound in cases presenting accelerated phases. In all these case, the EoS of k-essence increases with $a$ and run close to cosmological constant in the future. While the speed of sound $c^{2}_{\rm s}$ decreases with $a$ and runs close to zero in the future.

\begin{acknowledgments}
This study is supported in part by Research Fund for Doctoral
Programs of Hebei University No. 2009-155, and by Open Research
Topics Fund of Key Laboratory of Particle Astrophysics, Institute of
High Energy Physics, Chinese Academy of Sciences, No.
0529410T41-200901.
\end{acknowledgments}

\bibliography{apssamp}

\begin{thebibliography}{99}

\bibitem{Yang2010}
R.-J. Yang and S. N. Zhang, arXiv:0905.2683v3, to appear in MNRAS.

\bibitem{Arm00} C. Armend\'{a}riz-Pic\'{o}n, V. Mukhanov, and P. J. Steinhardt, Phys. Rev. Lett. \textbf{85}, 4438 (2000).

\bibitem{Damour1999} C. Armendariz-Picon, T. Damour, and V. Mukhanov, Phys. Lett. B \textbf{458}, 209 (1999).

\bibitem{Malquarti2003}
M. Malquarti, E. J. Copeland, A. R. Liddle, and M. Trodden, Phys. Rev. D \textbf{67}, 123503 (2003).

\bibitem{Aguirregabiria2004}
J. M. Aguirregabiria, L. P. Chimento, and R. Lazkoz, Phys. Rev. D \textbf{70}, 023509 (2004).

\bibitem{Abramo2006}
L. R. Abramo and N. Pinto-Neto, Phys. Rev. D \textbf{73}, 063522 (2006).

\bibitem{Rendall2006}
A. D. Rendall and Class. Quantum Grav. \textbf{23}, 1557 (2006).

\bibitem{Das2006}
R. Das, T. W. Kephart, and R. J. Scherrer, Phys. Rev. D \textbf{74}, 103515 (2006).

\bibitem{Chiba2009}
T. Chiba, S. Dutta, and Robert J. Scherrer, Phys. Rev. D \textbf{80}, 043517 (2009).

\bibitem{Cruz2009}
N. Cruz, P. F. Gonzalez-Diaz, A. Rozas-Fernandez, and G. Sanchez, Phys. Lett. B \textbf{679}, 293 (2009).

\bibitem{Granda2009}
L. N. Granda and A. Oliveros, arXiv:0901.0561v3

\bibitem{Gao2010}
X.-T. Gao and R.-J. Yang, Phys. Lett. B \textbf{687}, 99 (2010).

\bibitem{Arroja2010}
F. Arroja and M. Sasaki, Phys. Rev. D \textbf{81}, 107301 (2010).

\bibitem{Chimento2009}
L. P. Chimento, M. Forte, and M. G. Richarte, Phys. Rev. D \textbf{79}, 083527 (2009).


\bibitem{Sur2009}
S. Sur and S. Das, JCAP \textbf{0901}, 007 (2009).

\bibitem{Bilic2008}
N. Bilic, Phys. Rev. D \textbf{78}, 105012 (2008).


\bibitem{Scherrer2004}
R. J. Scherrer, Phys. Rev. Lett. \textbf{93}, 011301 (2004).

\bibitem{Daniele2007}
B. Daniele M. Sabino, and P. Massimo, Mod. Phys. Lett. A \textbf{22}, 2893 (2007).

\bibitem{Bose2009}
N. Bose and A. S. Majumdar, Phys. Rev. D \textbf{79}, 103517 (2009).



\bibitem{Yang2009} R.-J. Yang and X.-T. Gao, Chin. Phys. Lett. \textbf{26}, 089501 (2009).

\bibitem{Yang2008} R. J. Yang, S. N. Zhang, and Y. Liu, JCAP \textbf{0801} 017 (2008).

\bibitem{Yang2008a} R.-J. Yang and S. N. Zhang, Chin. Phys. Lett. \textbf{25} 344 (2008).


\bibitem{Sen2006}
A. A. Sen, JCAP \textbf{0603}, 010 (2006).

\bibitem{Putter2007}
R. de Putter and E. V. Linder, Astropart. Phys. \textbf{28}, 263 (2007).

\bibitem{Garriga1999}
J. Garriga and V. F. Mukhanov, Phys. Lett. B \textbf{458}, 219 (1999).

\bibitem{Unnikrishnan2010}
S. Unnikrishnan and L. Sriramkumar, Phys. Rev. D \textbf{81}, 103511 (2010).


\bibitem{Babichev2008}
E. Babichev, V. Mukhanov and A. Vikman, JHEP\textbf{0802}, 101 (2008).

\bibitem{Bruneton2007}
Jean-Philippe Bruneton, Phys. Rev. D \textbf{75}, 085013 (2007).


\bibitem{Kang2007}
J. U. Kang, V. Vanchurin, and S. Winitzki, Phys. Rev. D \textbf{76}, 083511 (2007).

\bibitem{Bonvin2006}
C. Bonvin, C. Caprini, and R. Durrer, Phys. Rev. Lett. \textbf{97}, 081303 (2006).

\bibitem{Gorini2008}

V. Gorini, A. Y. Kamenshchik, U. Moschella, O. F. Piattella, and A. A. Starobinsky, JCAP \textbf{0802}, 016 (2008)

\bibitem{Ellis2007}
G. Ellis, R. Maartens, and M. MacCallum, Gen. Rel. Grav. \textbf{39}, 1651 (2007).

\bibitem{Chim04}
L. P. Chimento, Phys. Rev. D \textbf{69}, 123517 (2004).

\end{thebibliography}

\end{document}